# Tools for Network Traffic Generation
# - A Quantitative Comparison

*Matthew Swann\*, Joseph Rose \*, Gueltoum Bendiab\*, Stavros Shiaeles \*, Nick Savage\**

*\*Cyber Security Research Group, University of Portsmouth, PO1 2UP, Portsmouth, UK*
Matthew.Swann@myport.ac.uk, Joseph.Rose@myport.ac.uk, gueltoum.bendiab@port.ac.uk,
sshiaeles@ieee.org, nick.savage@port.ac.uk

*Abstract*—Network traffic generators are invaluable tools that allow for applied experimentation to evaluate the performance of networks, infrastructure, and security controls, by modelling and simulating the communication packets and payloads that would be produced by machines and devices on the network. Specifically for security applications, these tools can be used to consistently simulate malicious activity on the network and test the components designed to detect and mitigate malicious activities, in a highly reliable and customisable way. However, despite the promising features, most of these tools have some problems that can undermine the correctness of experiments. The accuracy of the simulation results depends strongly on the performance and reliability of the used generator. Thus, in this paper, we investigate the performance and accuracy of three of the most reviewed network traffic generators in literature, namely Cisco TRex, Ostinato and Genesids. Mainly, the comparative experiments examine the strengths and limitations of these tools, which can help the research community to choose the most suitable one to assess the performance of their networks and security controls.

## I. Introduction

Network Traffic Generators (NTGs) are vital tools in the networking and security fields [1, 2]. They mainly focus on creating and injecting crafted network traffic into a network that can later be consumed by other devices in a controlled fashion. This is usually done to assess the performance of networks, infrastructures, security controls, and other aspects [1]. They can be implemented over both hardware and software platforms. Hardware platforms are more accurate and can achieve better performance, but they are expensive, closed source, and implemented on dedicated high-performance hard- ware [3]. Further, this kind of traffic generation tools is generally developed by firms and preconfigured to carry out a specific type of tests, and therefore difficult to customize [2]. Whereas, software-based generators are usually open-source, cheaper, but slower and less accurate [1, 3]. These tools are the most wildly used in the networking field mainly due to their high flexibility and open-source nature that allows for easy modifications and extensions based on specific research goals [1]. In the last years, a large number of software-based traffic generators have been proposed in the literature based on different methodologies [2], and most of them were adapted to the current need of network environments [3].

Despite the promising features provided by these tools, the results obtained by experiments are strictly dependent on the ability of the generators themselves to accurately emulate or replicate the network traffic pattern as it is requested by the operator [2, 3]. Thus, in this paper, we propose a quantitative evaluation of the most used open-source NTGs in the literature, namely Cisco TRex, Ostinato and Genesids. In particular, this paper, proposes a quantitative comparison of rate from a virtualised traffic generator to assess the performance, scalability and suitability of that NTG for replaying, or generation malicious traffic for testing, especially in security applications.

The remainder of this paper is organised as follows. Section 2 provides an overview of some prior work that is similar to our work herein. Section 3 presents the chosen network traffic generators Cisco TRex, Ostinato and Genesids, along with a high-level comparison between these three tools. In section 4, we present our testing methodology, the metrics used and an analysis of the obtained results. Finally, section 5 concludes the paper and outlines future work.

## II. RELATED WORK

In recent years, many research works have been studied in regard to the performance and precision of the software packet generator, for different purposes. In particular, Kolahi et all. [4] in 2011, proposed a quantitative comparison between four network traffic generators, namely Iperf, Netperf, D- ITG and IP Traffic. The comparison was done based on the throughput and payload size metrics. The experiments were conducted on both Windows and Linux operating systems. This study found that switching the performance monitoring tool used to collect data can make a quantifiable difference in the measured data throughput. Also, they found that most performance tools are designed to run best on Unix/Linux platforms, a factor that is important when considering the operating system of our packet generation system. In previous work, Avallone et al. [5] compared their NTG product with Mtools, Rude & Crude, MGEN, Iperf and UDP Generator. In this work, the experiments were performed on a Linux machine and only monitored the bandwidth of the link using the UDP protocol. Another interesting study proposed by Molnar et al [2] pro- posed a common set of metrics that can be used to facilitate the evaluation and also the comparison of different traffic genera- tors. The proposed metrics divided the studied metrics into five main categories, which are packet-based metrics, flow-based metrics, scaling characteristics and QoS/QoE related metrics. This study found that most of the research work proposed in this context lack any kind of quantitative comparisons and some relevant aspects of traffic characteristics (e.g., multi- scaling properties) are not considered in the evaluation of recent NTGs. The paper claimed that it is important that to validate the packet streams which are being produced by the traffic generators we choose, by capturing the packet streams with a packet sniffer such as Wireshark [6] or TCPdump [7]. This will increase the certainty that the chosen NTG will consistently generate expected traffic, and also help to check for malformed packets that could impact the validity of the testing.

In [8], Horn et al proposed an empirical comparison of four different implementations of NTGs that allowed for continuous generation of self-similar time series. Fundamentally this comparison focused on the accuracy of data and variance in results, which is not the metric we are using to compare our traffic generators on, but rather the rate of packets transmitted. However, we can apply the principles of repeatability and accuracy informed by the statistical analysis provided by the paper to our testing methodology. In a similar context, Emmerich et al [9] compared the suitability of different net- work traffic generators for network testing, by making specific comparisons between hardware, and the impact of CBR traffic on different CPU architectures and packet generators. They also investigated the impact of different I/O frameworks on the performance of the studied NTGs. Another work by A. Botta et al [5] has analysed the performance of the most used packet- level traffic generators in literature, MGEN, RUDE/CRUDE, TG and D-ITG. This study has pointed out the lack of accuracy of contemporary traffic generators- comparing the throughput of traffic generators such as Iperf [10], Mgen [11] and Rudecrude [12] to a calculated expected value. This is a great comparison, as it provides a frame of reference for the expected values of the traffic generators tested. The testing was conducted using a physical machine and Gigabit Ethernet connection, so, therefore, does not account for the difference in the performance of a virtualised solution.

In our review, we have found that there is no contemporary literature that covers the use cases for which the testing is done. Specifically, this is a quantitative comparison of rate from a virtualised traffic generator. This is to assess the performance, scalability, and suitability of the product for combination with the replay, or generation of malicious traffic for testing. Therefore, we can bring together aspects of previous work cited herein to inform and improve our methods of testing.

### III. Traffic Generation Tools

Traffic generators are used to create and then transmit crafted network traffic into a network, that can be then later utilised by other devices in such scenarios as IDS (Intrusion Detection Systems) alert generation and malicious network traffic simulation. They make use of a physical, high-level address and function, for all intents and purposes, they act as another device on the network. When set up, the traffic generator will either connect to the network and establish itself as a device or utilise the host that it is installed on, virtual or otherwise, to interact with the network on the same network interface. It will then proceed to generate newly generated packets, targeted at preconfigured devices on the network. Many traffic generators can also use the gateway to route packets so that it appears as though multiple other devices are connected to it and interacting with real devices on the network. The ensuing traffic generated is highly configurable and can be used to simulate many different network usage scenarios, for example:

- Testing if a network can handle the implementation of a new, network-intensive application.
- Stress testing a network, to see whether it can withstand DOS attacks from internal and external sources.
- Testing IDS Systems with threat signatures coming from within the network.

The behaviour of the traffic generated, what layer it functions on, the configuration of different payloads, headers and flags etc. varies from product to product. Some can be based on recordings of traffic from the real world, whereas others will manually craft streams of packets, or alternatively use random data built using an algorithm in the generator. In order to create and demonstrate realistic attacks on the Cyber-Trust network, the following solutions were considered and tested.

### A. Cisco TRex

The TRex [13] traffic generator is an open-source, flexible, traffic generator that can be used in a stateless and stateful configuration. When being used statelessly, it is possible to craft multiple packet generation streams and alter any field within the packet, including headers, trailers, payload, flags- even sending deliberately corrupt packets to test the network's response. These settings are passed through an algorithm which crafts the packets, assigns them a packet ID and transmits them across the network. TRex is a good stress testing tool and can be used for testing firewalls, load balancing and other load-based tasks as it can concurrently construct packets in multiple streams. However, it does not support routing emulation, making its usage for simulating multiple devices, without multiple instances, difficult.

### B. Ostinato

Ostinato [14] is a similar, popular network traffic and packet generator. Like with TRex you can create your own packet generation streams; you can also configure them in an in- depth manner, with considerations towards stream rates, bursts, and the number of packets generated. This is a tool suited to network load testing and functionality testing. It also has support for an extremely wide array of protocols, to allow you to tailor packets and stack protocols however you wish. This can be extremely useful for testing edge-case errors while still having extremely high-performance. Ostinato is stateless and as such doesn't support connection-oriented setups, e.g., a headless browser interaction with a website, which can make it limited for security testing.

### C. Genesids

Genesids [15] is a traffic generator specifically designed for high-volume, tailored, security testing of IDS systems. Often to achieve this task, real-world captured traffic is used and replayed across the network. However, this is neither adaptable or suitable for use on a large scale. Genesids allows the specification and generation of application layer payloads, using definitions from Snort rules to craft its traffic. It's ideal for use alongside another traffic generator in order for it to produce realistic background traffic and to maintain the typical load pattern of a network. due to this it's able to reliably produce a large variety of malicious traffic repeatably.

## IV. Testing Methodology

The testing approach has been proposed in the context of the Cyber-Trust project (https://cyber-trust.eu/) in order to evaluate the attack coverage by the Cyber-Trust components and their ability to identify attacks precisely. The traffic generators are used to create an adaptable, realistic attack scenario that can be easily repeated and adapted in a controlled way to allow for thorough and repeatable testing of the Cyber-Trust com- ponents. The Cyber-Trust network consists of NIDS systems running deep and machine learning techniques on Game- theoretic framework in order to provide recommendations for IoT service providers based on the information regarding vulnerabilities. The premise of the test for each generator was to generate and transmit as many TCP packets as possible within an allocated time period of one minute. This would provide an objective metric that shows the capability and receptiveness to scale of each solution. Each generator was deployed inside a virtual environment that running 2 processors (CPU cores) with an execution cap of 100% and nested paging enabled, with 4GB of RAM and 10GB of disk space (located in a fixed VDI). While the tests have been conducted in a virtualised environment, which will inevitably lower performance com- pared to hardware-based network infrastructure, the purpose of this testing was to benchmark the performance and suitability of each generator for deployment and use in the Cyber-Trust network for IDS and malicious network traffic testing, which is situated within a virtualised network.

To provide as fair of a comparison as possible, each generator was set to allow the maximum transmission rate allowed by the virtualised hardware. As TRex is by default almost exclusively hardware limited, the only modifications to the TRex configuration were modifying the default TCP traffic template and changing the distribution to "seq" (sequential) and the "tcp ageing" factor to 1. In Ostinato, removing the usual software limit of the host running drone, to generate traffic at the maximum hardware allowed speed, is done by configuring a stream with "packets/Sec" set to 0 in "continuous" mode. As Genesids focus is customisable malicious traffic generation, rather than capacity, it does not enforce rate limiting by default, only as an option, and as such did not require adjustment. As Genesids does not provide the ability to specify a packet size, rather generating a specific type of packet according to provided Snort rules, the packet size was kept default for each solution.

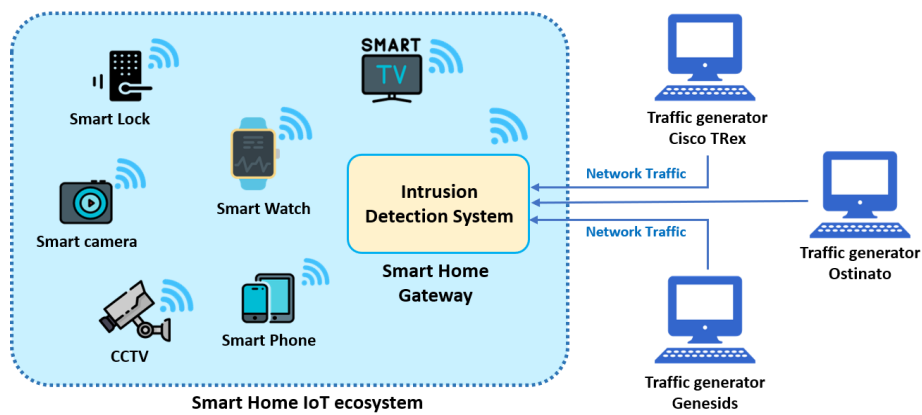

Fig. 1. High-level overview of the Testbed

### A. Software Unique Configuration

It is worth noting that software specific configuration changes were made from the default settings. TRex was left mostly default, however Ostinato requires specification of the layer 1-4 traffic, which was as follows: L1: Mac, L2: Ethernet II, L3: IPv4, L4: TCP and no special signatures, flags, or override headers. Genesids requires a list of Snort rules to generate packets from. The rules used in testing were the Snort3 Community Rules v3.0. It is also worth noting that, despite being under TRex's recommended usage specification, the low-performance, and low-footprint modes [16] have not been enabled in our testing.

### B. Data Collection

The data for each test was collected by utilizing TCPdump to generate a PCAP file from the captured traffic, allowing viewing of the generated packets from the host. These were captured with the purpose of analysing the packet contents, in order to validate that the payloads had been set correctly, and to check for malformed or corrupt packets during traffic generation. The resource usage of each NTG was collected using a short bash script, writing the output of top [17] specified for the process ID of the NTG to a file. This allows us to collect the CPU and RAM usage at intervals of one second, over the period of one minute- matching our data collection. This was done to provide accurate system monitoring, without eating into the resources available to the virtual machine.

## V. Results Analysis and Discussion

### A. Cisco TRex

The figures below (Fig. 2 and Fig. 3) show the resource usage of TRex. Observing the network traffic produced by TRex, the general trend shows a slow start, with the rate of packet generation increasing, then increasing to a baseline of 10929 Packets/s at 8 seconds, before increasing rapidly to 104180 between the 9 and 10 second mark. The rate then becomes less variable, hovering around 110000 Packets/s, with two noticeable spikes to 140000 Packets/s occurring consecutively at the 50 and 53 second marks, where the rate peaks at 140141 Packets/s before normalising. This yields an average rate of 98810 Packets/s generated and transmitted, with a standard deviation of 37526 Packets/s.

As shown in Fig. 3, the CPU usage of TRex is high, with the lowest utilisation of 87.2%, occurring at the 500KB/s generation test. Then, the usage continually increases by around 1-2% per 500KB/s, with small but consistent drops in utilisation occurring over the minute period. It is worth noting that when TRex utilises 100% of the first CPU core, it immediately begins using the second CPU core- which is represented on the graph as the points at over 100% CPU use. However, the memory usage (see Fig. 2) shows a consistent pattern. The utilisation remains almost entirely static for the duration of each test- increasing by around 0.18% per 500KB increase in throughput. This is a fairly linear increase in memory usage- with the utilisation staying mostly constant, apart from one jump from 3.0% to 3.1% at the 58 second mark.

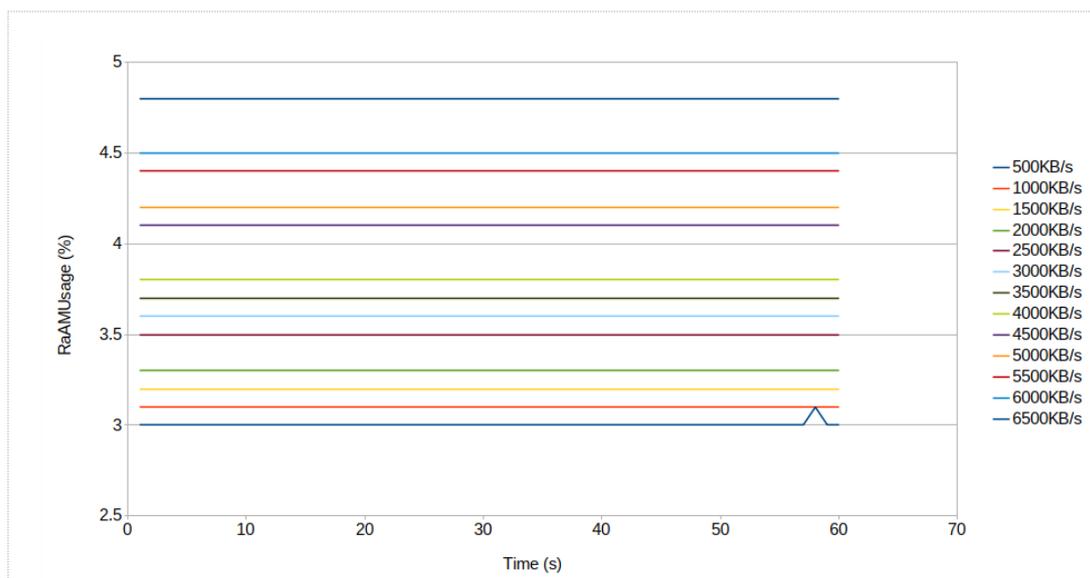

Fig. 2. A graph to show the RAM usage of Cisco TRex at different bandwidths

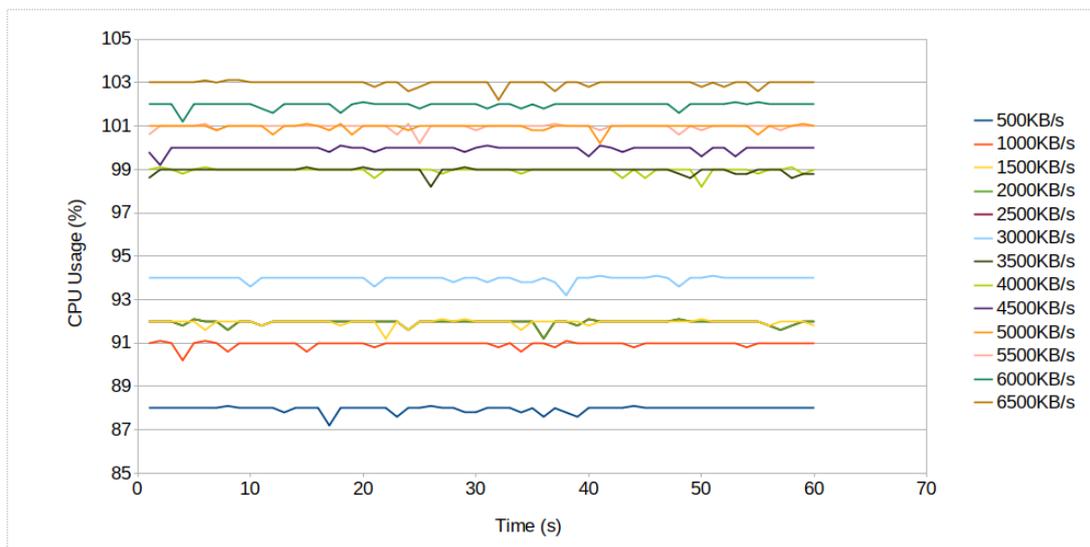

Fig. 3. A graph to show the CPU usage of Cisco TRex at different bandwidths

### B. Ostinato

In comparison, the Ostinato packet generation starts much higher, at 121369 Packets/s, with a sharp increase to 167818 at the 5 second mark, before rapidly decreasing again. There is a consistent, large, and often extreme variance within this rate, most notably in the period between 18 and 25 seconds, during which the rate drops from the peak at 188608 Packets/s to its second lowest at 89675, a drop of 47.55%. Ultimately, this data shows an average rate of 127271 Packets/s generated and transmitted, with a standard deviation of 23887 Packets/s.

Ostinato's CPU utilisation (see Fig. 5) increases by around 2-3% with each increase of 500KB/s generated. There are small but frequent dips, varying from 0.5% to up to 2% in utilisation, occurring at varying time intervals throughout each usage test. The increase in CPU utilisation per 500KB/s load is, just like TRex, extremely regular- with increases in the range of 2% and 4% for each increase. Similarly, to TRex, the RAM usage for each test, shown in Fig. 4, is extremely consistent, with very little variation occurring over the minute period tested for each throughput value tested. Increases in memory usage were observed on throughput increases from 3000KB/s to 3500KB/s, 4500KB/s to 5000KB/s and 6000KB/s to 6500KB/s. This low memory footprint is a contributing factor to Ostinato's good performance in our low-resource virtualised testing environment, with a maximum memory usage of just 0.6% during Ostinato's highest throughput tested-7500KB/s.

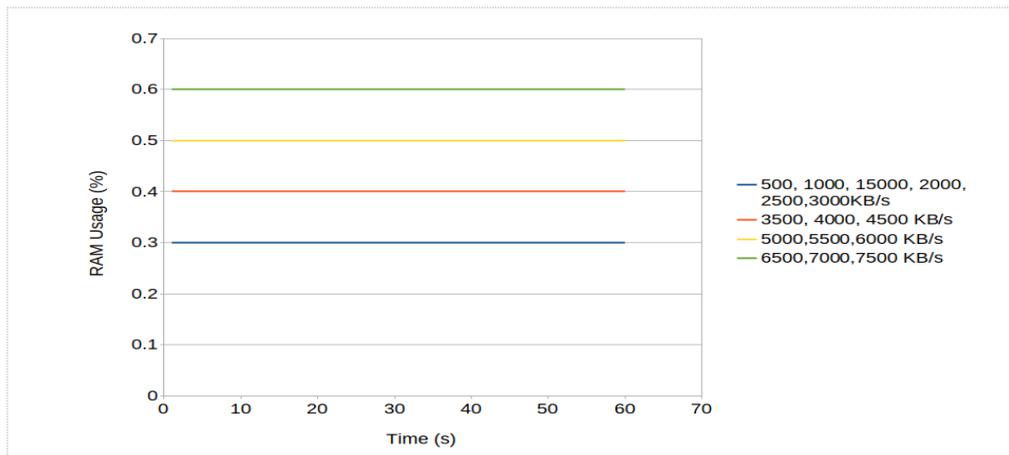

Fig. 4. A graph to show the RAM usage of Ostinato at different bandwidths

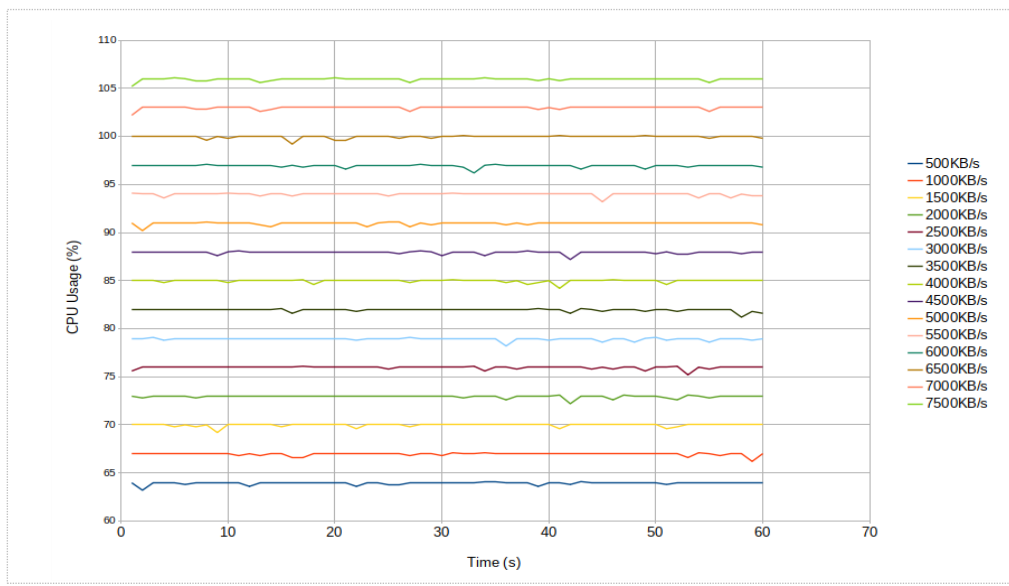

Fig. 5. A graph to show the CPU usage of Ostinato at different bandwidths

### C. Genesids

Genesids provides a much lower but much more stable rate of packet generation and transmission. The rate starts at 11552 Packets/s and the overall trend stays consistent around this baseline, with the average rate of being 11021 Packets/s, with a standard deviation of 844 Packets/s. As illustrated in Fig. 7, in contrast to the other tests, who both used 100% of the first core available, Genesids has a much lower overall CPU usage, peaking at just 29.1% utilisation. However, it does experience a much more frequent and much larger variance in usage, with a maximum variance of 4.1%. It differs in its memory usage, consuming a much larger percentage of RAM, with the trend increasing the longer each test continues. these results in maximum memory usage of 29% (see Fig. 6).

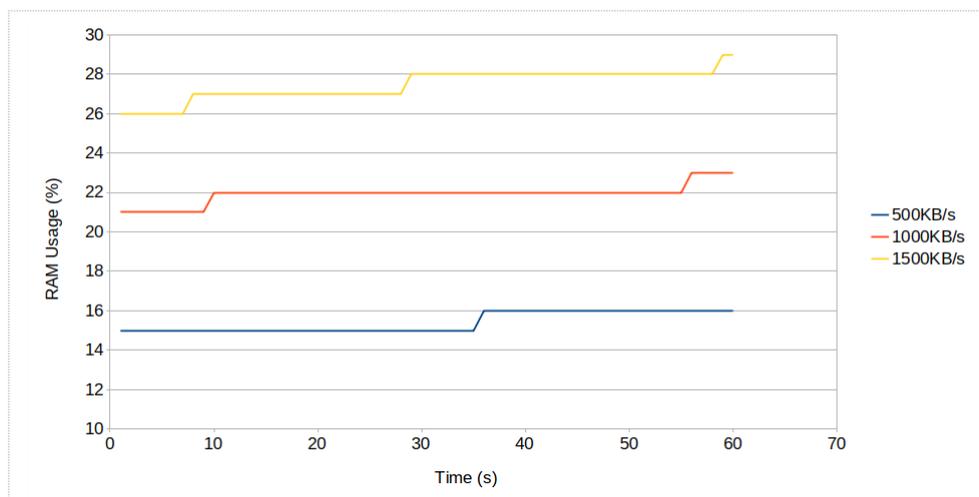

Fig. 6. A graph to show the RAM usage of Genesids at different bandwidths

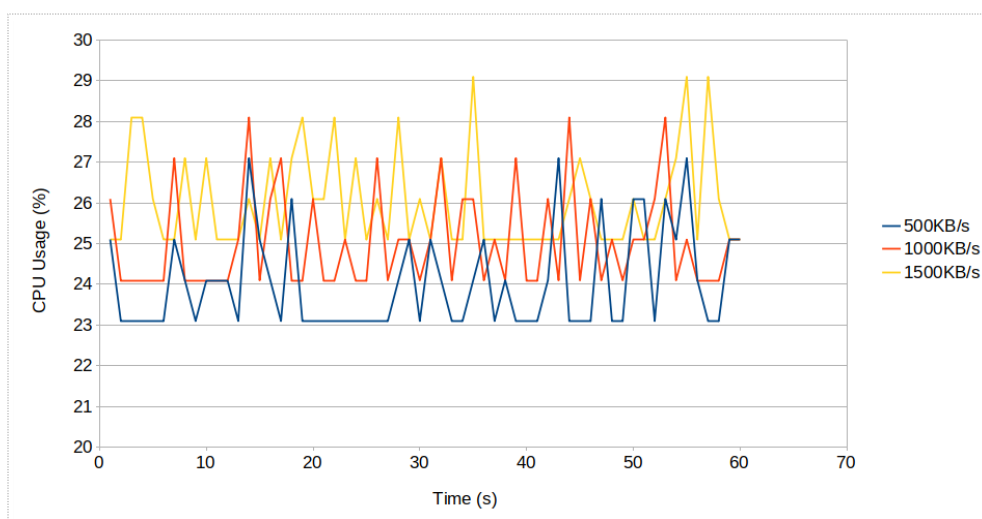

Fig. 7. A graph to show the CPU usage of Genesids at different bandwidths

### D. Analysis

Ultimately our results show that, for our host configuration, Ostinato produced by far the greatest level of traffic, producing on average 28460 Packets/s than the network traffic generator with the second greatest rate of packet generation and trans- mission, TRex. Furthermore, while comparing the data for the one minute time allocation of testing, it shows that Genesids had the least variance, at 7.65%, followed by Ostinato at 18.77% with TRex having the most variance at 37.98%. However, TRex's standard deviation is not representative of the mostly stable trend observed. Calculating TRex's standard deviation after 9 seconds, which excludes the time it takes TRex's rate to reach stability from the beginning of the test, the variance is brought down to 9.95%.

### E. Evaluation

Although these results do not definitively prove that Ostinato is the best packet generator out of the three compare traffic generators. It does provide an objective metric to state that, for this low resource, virtualised configuration- Ostinato performed the best for the metric of highest average Packets/s generated while providing an infrequent, but notable variance of 18.77%. The architecture of each packet generator is likely the explanation for the varied result, as each solution uses a different architecture and generation algorithm. Ostinato uses a unique architecture written in C$^{++}$ to transmit packets, where Genesids utilises libcurl [18]. TRex uses a different approach, utilising the DPDK Kernel interface [19] to directly interact with the Linux kernel, behaving as a loadable kernel module to interface directly with the kernel network stack. This approach may function comparatively worse on a single-threaded application like the one tested- however it may be more scalable with more allocated processors. This low resource allocation could explain the large divide between TRex, Genesids and Ostinato. For example, The Enterprise Stack, as well as TRex's own hardware recommendations [20] per their installation manual [16], recommend that a virtual machine running TRex should have at least 4GB of Memory and 4 virtual processors.

In our testing, we only allocated two. Further testing would be necessary to ascertain how much of a difference this makes.

## VI. Conclusion

In this paper, we presented our testing methodology of the traffic generators Cisco TRex, Ostinato and Genesids. The implantation of the packet generators is done within the Cyber- Trust network to test the IDS and Machine Learning com- ponents to a measurable and repeatable level in a controlled environment. By combining the clean packet generation from the high-performance Ostinato with the customisable malicious packet generation ability of Genesids; we can leverage a robust, scalable performance of both solutions to create adaptable and realistic attack scenarios. The scenario of attack can be easily repeated and adapted in a controlled network environment. Furthermore, we can adapt the malicious and clean packet generation to simulate attacks on the whole network, key devices, or individual devices as well as inside, or outside of the Smart Home network. This will allow for thorough and repeatable testing of components within a virtual environment. This comes with many benefits, a large part of the simulation of rules and network traffic to the components running IDS systems with integrated machine learning components can allow for the automation of training and measurements in detection accuracy and false-positive rates of many components, all without the need for manual exploitation or attacks of networks which can be costly and time-consuming. This can be invaluable for the research community as a whole when dealing with the limitations, requirements and availability of many different and emerging packet generators that can aid in the collation of needed information. The accuracy of our testing can be evidenced through the consistency of the results we acquired throughout our testing, with an average standard deviation for repeat testing of ±21.29% for our NTG performance tests. We can also compare our results to that of known benchmarks. Comparing our TRex results with that of the TRex virtual machine installation and testing guide [16] a result of "9.99 Kpkt/sec" is shown, a result mirrored by our average of 9.64Kpkt/sec despite the decreased resources.

While there are still restrictions in hardware requirement, the limitations herein are held against the availability of network hardware resources and can be scaled as seen fit for purposes of your own. As such, the benefits of our testing can be seen. The tests were all conducted in the same environment with constant machine specifications and targets. This resulted in each NTG having the same opportunity to perform to its best, under the deliberately constructed poor conditions. This tested a factor that is not often considered with high- performance, multi-state NTG's- which is their scalability to low-end, often virtualised, use cases. Ultimately, our research and testing could be expanded by comparing the resource usage and performance of the NTG's over a longer time period to explore performance for long-term use.

### ACKNOWLEDGMENT

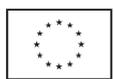 This project has received funding from the Euro- pean Union's Horizon 2020 research and innovation programme under grant agreement no. 786698. This work reflects authors' view and Agency is not responsible for any use that may be made of the information it contains.